\documentclass{amsart}
\usepackage{amsmath, color}
\usepackage{amsfonts}
\usepackage{latexsym}
\usepackage{ifpdf}
\usepackage{graphicx,amssymb,lineno}
\usepackage{amssymb}
\usepackage{mathrsfs}
\usepackage{cite}
\usepackage{extarrows}
\usepackage{ulem}
\usepackage{placeins} 
\usepackage{booktabs}
\usepackage{array}
\usepackage{tikz}
\usepackage{tabularx}
\usepackage{booktabs}   
\usepackage{pifont}     
\usepackage{ytableau}
\usepackage{ytableau}
\usepackage{bm}

\ytableausetup{
  mathmode,
  boxsize=1.3em,
  smalltableaux
}

\usetikzlibrary{arrows.meta,positioning,calc,decorations.pathreplacing}

\allowdisplaybreaks[4]
\newcommand{\Var}{\text{Var}}
\newcommand{\Tr}{\operatorname{Tr}}

\newtheorem{theorem}{Theorem}[section]

\newtheorem{proposition}[theorem]{Proposition}

\theoremstyle{definition}

\theoremstyle{remark}

\numberwithin{equation}{section} \errorcontextlines=0
\begin{document}

\title[Sharp State-independent uncertainty relations for Multipartite Systems]
{Sharp State-independent uncertainty relations for Multipartite Systems}
\author{Yiling Wang}
\address{School of Mathematics,
South China University of Technology,
   Guangzhou 510640, ChinaA}
\email{ywang327@scut.edu.cn}

\author{Naihuan Jing}
\address{Department of Mathematics,
   North Carolina State University,
   Raleigh, NC 27695, USA}
\email{jing@ncsu.edu}

\thanks{*Corresponding author: Naihuan Jing}

\begin{abstract}
Uncertainty relations constrain the fluctuations of incompatible observables, but most familiar bounds depend on the quantum state.
State-independent uncertainty relations instead ask how much fluctuation remains unavoidable for every quantum state. For observables generated by a continuous symmetry, sharp state-independent bounds are known when the symmetry representation is irreducible. Multipartite collective systems, however, generally appear as reducible tensor-product representations of a symmetry algebra, which raises the question of how to determine their total uncertainty.

We resolve this problem for multipartite quantum systems with a compact semisimple symmetry algebra $\mathfrak g$. 
Exploiting the symmetry structure, we formulate a general framework for state-independent uncertainty based on representation theory.
The total 
variance admits an exact decomposition into intrinsic fluctuations within irreducible sectors and a nonnegative dispersion between sectors.
This yields the sharp state-independent bound for the total variance $\Delta_\rho^2(\mathfrak g)$ on the multipartite Hilbert space $\mathcal H$
\[
\min_{\rho}\Delta_\rho^2(\mathfrak g)
=
\min_{\lambda\in\Lambda(\mathcal H)}
2\langle\lambda,\delta\rangle,
\]
where $\rho$ is any density operator on $\mathcal H$, $\Lambda(\mathcal H)$ is the set of highest weights $\lambda$ labeling 
those sectors, and $\delta$ is the Weyl vector. This demonstrates that the ultimate uncertainty is completely controlled by its intrinsic symmetry structure. 
As an application of the general theory, we confirm our previous conjecture [JPA 59, 155302 (2026)] that the total uncertainty floor of collective spin-$1/2$ systems depends only on the parity of the particle number. 
We further illustrate the framework for multipartite spin-$1$ systems, demonstrating that the same symmetry-sector mechanism persists beyond spin-$1/2$.
\end{abstract}
\maketitle
\section{Introduction}
\label{sec:introduction}

Since the foundational formulations of quantum uncertainty
\cite{heisenberg1927,kennard1927,robertson1929,schrodinger1930},
understanding the constraints that incompatible observables impose on
quantum fluctuations has remained a central problem in quantum theory. Conventional uncertainty bounds are typically state dependent and may
become trivial for particular states even when the observables remain
incompatible \cite{maccone2014}. This motivates a
complementary question: how much uncertainty is unavoidable for
every state of a given quantum system?  State-independent
uncertainty relations address precisely this global question by seeking
universal bounds that hold uniformly over the entire state space
\cite{dammeier2015,schwonnek2017,deGuise2018}.  Beyond their foundational significance, uncertainty relations play
important roles in spin squeezing, entanglement characterization, and
quantum metrology
\cite{kitagawa1993,GuhneToth2009,giovannetti2011,pezze2018}.
More recently, they have been connected quantitatively to quantum Fisher
information, entanglement monotones, and achievable precision in
multiparameter estimation
\cite{DuLiuFadelVitaglianoHe2026}; see also Ref.~\cite{Xiao2026} for a
recent perspective.

Continuous symmetries provide a particularly structured setting in
which the state-independent problem can be addressed through
representation theory.  Let $\mathfrak g$ be a compact semisimple Lie
algebra acting unitarily on a finite-dimensional Hilbert space
$\mathcal H$, and let $\{X_a\}$ be an orthonormal basis of $\mathfrak g$ consisting of the Hermitian operators.  For a quantum state $\rho$,
we consider the total generator variance
\begin{equation}
\Delta_\rho^2(\mathfrak g)
=
\sum_a
\left[
\Tr(\rho X_a^2)
-
\bigl(\Tr(\rho X_a)\bigr)^2
\right],
\label{eq:intro-total-variance}
\end{equation}
which measures the total fluctuation of the symmetry generators. This definition is independent from the choice of orthonormal generator basis $\{X_a\}$.

When the symmetry $\mathfrak g$ acts irreducibly on a highest-weight representation
$V(\lambda)$, indexed by the highest weight $\lambda$, the total uncertainty admits the sharp state-independent bound
\begin{equation}
\Delta_\rho^2(\mathfrak g)
\geq
c(\lambda)
=
2\langle\lambda,\delta\rangle ,
\label{eq:intro-irrep-bound}
\end{equation}
where $\delta$ is the Weyl vector
\cite{deGuise2018,HuJingZhangXiao2026}.
The bound is attainable on generalized coherent states associated with
extremal-weight orbits
\cite{Delbourgo1977,DelbourgoFox1977,
ZhangFengGilmore1990,Perelomov1986}.
Thus, for an irreducible representation, the symmetry label itself
determines a sharp intrinsic uncertainty level.

The irreducible setting, however, is not generic for collective quantum
systems.  If each of $n$ constituents carries a single-particle
representation $V$, the collective Hilbert space
$\mathcal H_n=V^{\otimes n}$ generally forms a reducible representation
containing several irreducible symmetry sectors, often with nontrivial
multiplicities
\cite{FultonHarris,Hall2015,Hayashi2017}. Such representation structures occur widely in quantum information:
Clebsch--Gordan and Schur decompositions organize collective degrees of
freedom \cite{BaconChuangHarrow2006}, while closely related structures
underlie reference-frame and superselection constraints
\cite{bartlett2007} and decoherence-free or noiseless encodings
\cite{lidar1998,KnillLaflammeViola2000}.
Reducibility is therefore not merely a formal generalization of the
irreducible problem, but an intrinsic feature of many-body symmetry.

Once several irreducible sectors coexist, the bounds of the sectors alone no longer determine the uncertainty of a general state with support
across different sectors. How are the intrinsic uncertainties of the sectors combined, and does their coexistence generate fluctuations
beyond a weighted average of the sectorwise contributions? Which accessible sector ultimately fixes the sharp state-independent floor of the full reducible representation? A concrete indication of such a mechanism appears in collective spin-$1/2$ systems, where a recent
analysis identified an odd--even structural pattern: odd particle numbers exclude the trivial representation, whereas even particle
numbers admit singlet sectors capable of supporting vanishing collective variance \cite{Wang2026}. 
This parity dependence suggests a more general representation-theoretic mechanism, but identifying it requires a state-independent theory formulated directly on reducible representation spaces.

To place the present problem in context,
Table~\ref{tab:SIUR-comparison} summarizes representative
state-independent variance-based uncertainty relations most closely
related to the present work, with particular emphasis on multipartite
systems and the underlying representation structure.

\begin{table}[h]
\centering

\caption{
Comparison of representative state-independent variance-based
uncertainty relations relevant to the present framework.
A check mark (\ding{51}) indicates that the corresponding feature is
treated explicitly, while a dash indicates that it is not the primary
focus of the cited work. ``General $\mathfrak g$'' refers to a
formulation for general compact semisimple symmetry algebras. The asterisk indicates the specific reducible collective
$\mathfrak{su}(2)$ representations arising in the multipartite
spin-$1/2$ setting of Ref.~\cite{Wang2026}.
}
\label{tab:SIUR-comparison}

\small
\renewcommand{\arraystretch}{1.25}

\begin{tabular}{
@{}
>{\raggedright\arraybackslash}p{0.25\linewidth}
>{\centering\arraybackslash}p{0.17\linewidth}
>{\centering\arraybackslash}p{0.17\linewidth}
>{\centering\arraybackslash}p{0.17\linewidth}
>{\centering\arraybackslash}p{0.16\linewidth}
@{}
}
\toprule

\textbf{Reference}
&
\textbf{Multipartite}
&
\textbf{Reducible reps.}
&
\textbf{General $\mathfrak g$}
&
\textbf{Sharp bound}
\\

\midrule

Dammeier~et~al.~\cite{dammeier2015}
&
---
&
---
&
---
&
\ding{51}
\\[3pt]

Schwonnek~et~al.~\cite{schwonnek2017}
&
---
&
---
&
---
&
\ding{51}
\\[3pt]

de~Guise~et~al.~\cite{deGuise2018}
&
---
&
---
&
\ding{51}
&
\ding{51}
\\[3pt]

Wang~et~al.~\cite{Wang2026}
&
\ding{51}
&
\ding{51}$^{\ast}$
&
---
&
---
\\[3pt]

\textbf{This work}
&
\textbf{\ding{51}}
&
\textbf{\ding{51}}
&
\textbf{\ding{51}}
&
\textbf{\ding{51}}
\\

\bottomrule
\end{tabular}

\end{table}

The comparison highlights the distinction between two levels of
generality. State-independent variance bounds are available in several
settings, and sharp Lie-algebraic bounds are known for irreducible
representations. Multipartite spin-$1/2$ systems further demonstrate
that reducible representation structure can directly affect the
uncertainty floor. What remains to be established is a general
state-independent framework that treats reducible tensor-product
representations of compact semisimple symmetry algebras and identifies
how their irreducible sectors combine to determine the sharp global
uncertainty bound.

In this paper, we develop such a theory by exploiting the isotypic decomposition of the symmetry representation.
Resolving the Hilbert space into its irreducible symmetry sectors, we show that the total generator variance decomposes exactly into two contributions: intrinsic quantum
fluctuations within the individual sectors and a nonnegative intersector dispersion of their mean generator vectors, providing a symmetry-resolved quantum analog of the law of total variance.
Combining this exact decomposition with Eq.~\eqref{eq:intro-irrep-bound}, we obtain a sharp sector-resolved uncertainty relation and the exact global minimum
\begin{equation}
C_{\min}(\mathcal H)
\equiv
\min_{\rho}\Delta_\rho^2(\mathfrak g)
=
\min_{\lambda\in\Lambda(\mathcal H)}
2\langle\lambda,\delta\rangle,
\label{eq:intro-main-result}
\end{equation}
where $\Lambda(\mathcal H)$ denotes the set of highest weights labeling
the irreducible sectors contained in $\mathcal H$. The optimization over
the full quantum state space therefore collapses to a minimization over
the discrete set of accessible symmetry sectors.

The result identifies the lowest accessible irreducible sector as the
one that sets the state-independent uncertainty floor. For collective
spin-$1/2$ systems, this mechanism rigorously proved the conjectured
parity effect: odd particle numbers contain no singlet sector and hence
retain $C_{\min}=1/2$, whereas even particle numbers admit a singlet and
allow $C_{\min}=0$. Removing or restricting accessible symmetry sectors
shifts the sharp floor accordingly, while higher-spin tensor products
exhibit the same mechanism in the presence of nontrivial
multiplicities. Representation content, rather than Hilbert-space size,
therefore emerges as the organizing principle for state-independent
uncertainty in reducible collective systems.

\section{Exact variance decomposition and sharp uncertainty bound}

Collective observables generated by continuous symmetries naturally act
on reducible representations.  The central challenge in extending
state-independent uncertainty relations beyond irreducible systems is
therefore to determine how uncertainty contributions from distinct
symmetry sectors combine.  We show that this problem admits an exact
solution: the total variance separates into intrinsic quantum
fluctuations within each irreducible sector and a positive dispersion
associated with the coexistence of different sectors.

Consider a finite-dimensional unitary representation $V$ of a compact
semisimple Lie algebra $\mathfrak g$. The multipartite Hilbert space
$\mathcal H_n=V^{\otimes n}$ becomes a representation of $\mathfrak g$
through the coproduct action
\begin{equation}
x
\mapsto
\sum_{i=1}^{n}
I^{\otimes(i-1)}
\otimes x
\otimes
I^{\otimes(n-i)},
\qquad
x\in\mathfrak g .
\label{eq:coproduct-action}
\end{equation}
The resulting tensor-product representation admits the following
isotypic decomposition.

\begin{theorem}[Isotypic decomposition]
\label{thm:isotypic-decomposition}
Let $V$ be a finite-dimensional unitary representation of a compact
semisimple Lie algebra $\mathfrak g$, and let
$\mathcal H_n=V^{\otimes n}$. Then
\begin{equation}
\mathcal H_n
\cong
\bigoplus_{\lambda\in\Lambda_n(V)}
\left(
\mathcal M_\lambda\otimes V(\lambda)
\right),
\label{eq:prl-isotypic}
\end{equation}
where $\Lambda_n(V)$ is the set of highest weights labeling the
inequivalent irreducible representations appearing in $V^{\otimes n}$,
$V(\lambda)$ is the irreducible highest-weight representation labeled
by $\lambda$, and
\begin{equation}
\mathcal M_\lambda
\equiv
\operatorname{Hom}_{\mathfrak g}
\bigl(V(\lambda),\mathcal H_n\bigr)
\label{eq:multiplicity-space}
\end{equation}
is the corresponding multiplicity space.
\end{theorem}

\begin{proof}
By Weyl's complete reducibility theorem, the finite-dimensional
$\mathfrak g$-module $\mathcal H_n$ decomposes as
\begin{equation}
\mathcal H_n
\cong
\bigoplus_{\lambda\in\Lambda_n(V)}
V(\lambda)^{\oplus m_\lambda},
\label{eq:complete-reducibility}
\end{equation}
where $m_\lambda$ is the multiplicity of $V(\lambda)$ in
$\mathcal H_n$.

For two irreducible representations $V(\lambda)$ and $V(\mu)$,
Schur's lemma gives
\begin{equation}
\operatorname{Hom}_{\mathfrak g}
\bigl(V(\lambda),V(\mu)\bigr)
\cong
\begin{cases}
\mathbb C, & \lambda=\mu,\\
\{0\}, & \lambda\neq\mu.
\end{cases}
\label{eq:schur-hom}
\end{equation}
Using the additivity of $\operatorname{Hom}_{\mathfrak g}$ over direct
sums, Eq.~\eqref{eq:complete-reducibility} therefore yields
\begin{align}
\mathcal M_\lambda
=
\operatorname{Hom}_{\mathfrak g}
\bigl(V(\lambda),\mathcal H_n\bigr)
&\cong
\bigoplus_{\mu\in\Lambda_n(V)}
\operatorname{Hom}_{\mathfrak g}
\bigl(V(\lambda),V(\mu)^{\oplus m_\mu}\bigr)
\nonumber\\
&\cong
\operatorname{Hom}_{\mathfrak g}
\bigl(V(\lambda),V(\lambda)^{\oplus m_\lambda}\bigr)
\nonumber\\
&\cong
\mathbb C^{m_\lambda}.
\label{eq:multiplicity-hom}
\end{align}
Hence
\begin{equation}
\dim\mathcal M_\lambda=m_\lambda.
\end{equation}

The canonical evaluation map
\begin{equation}
\mathcal M_\lambda\otimes V(\lambda)
\longrightarrow
\mathcal H_n,
\qquad
T\otimes v\longmapsto T(v),
\label{eq:evaluation-map}
\end{equation}
is a $\mathfrak g$-equivariant map whose image is precisely the
$\lambda$th isotypic component of $\mathcal H_n$. Since
$\dim\mathcal M_\lambda=m_\lambda$, this identifies that component as
\begin{equation}
V(\lambda)^{\oplus m_\lambda}
\cong
\mathcal M_\lambda\otimes V(\lambda).
\end{equation}
Applying this identification to every irreducible type appearing in
Eq.~\eqref{eq:complete-reducibility} gives
\begin{equation}
\mathcal H_n
\cong
\bigoplus_{\lambda\in\Lambda_n(V)}
\left(
\mathcal M_\lambda\otimes V(\lambda)
\right),
\end{equation}
which proves Eq.~\eqref{eq:prl-isotypic}.
\end{proof}

See Refs.~\cite{FultonHarris,Hall2015,Hayashi2017} for the general
representation-theoretic framework.

The multiplicities appearing in the isotypic decomposition can also be computed directly from characters. Let $G$ be a compact connected Lie group with Lie algebra $\mathfrak g$, and let $\chi_V$ and $\chi_\lambda$ denote the characters of $V$ and $V(\lambda)$, respectively. Since characters are multiplicative under tensor products, 
\begin{equation} 
\chi_{V^{\otimes n}}(g)=
\chi_V(g)^n . 
\label{eq:tensor-character-general} \end{equation} By the orthogonality of irreducible characters, the multiplicity of $V(\lambda)$ in $V^{\otimes n}$ is therefore \begin{equation} 
m_\lambda=
\dim \mathcal M_\lambda=
\int_G
\overline{\chi_\lambda(g)}
\chi_V(g)^n
d\mu(g),
\label{eq:multiplicity-character}
\end{equation}
where $d\mu$ denotes the normalized Haar measure on $G$.

Of particular relevance below is the trivial representation $V(0)$. Since its character is $\chi_0(g)=1$, Eq.~\eqref{eq:multiplicity-character} reduces to 
\begin{equation} 
m_0 =
\dim \mathcal M_0=
\int_G
\chi_V(g)^n
d\mu(g),
\label{eq:singlet-multiplicity-general}
\end{equation}
which is the number of copies of the trivial
representation in $V^{\otimes n}$. The collective spin-$1/2$ and spin-$1$
systems considered below provide explicit examples of this general
character-theoretic multiplicity formula.

Since $\mathfrak g$ acts trivially on the multiplicity spaces
$\mathcal M_\lambda$, the collective generators preserve the isotypic
decomposition and take the block-diagonal form
\begin{equation}
X_a
=
\bigoplus_{\lambda\in\Lambda_n(V)}
\left(
I_{\mathcal M_\lambda}\otimes X_a^{(\lambda)}
\right),
\label{eq:prl-block-action}
\end{equation}
where $X_a^{(\lambda)}$ denotes the representation of the generator
$X_a$ on the irreducible factor $V(\lambda)$.

For a quantum state $\rho$ on $\mathcal H_n$, let $P_\lambda$ denote
the orthogonal projector onto the $\lambda$th isotypic component and
define the sector weight
\begin{equation}
p_\lambda
=
\Tr(P_\lambda\rho).
\label{eq:prl-sector-weight}
\end{equation}
For $p_\lambda>0$, the corresponding conditional state on the
irreducible factor $V(\lambda)$ is
\begin{equation}
\rho_\lambda
=
\frac{
\Tr_{\mathcal M_\lambda}
(P_\lambda\rho P_\lambda)
}{p_\lambda}.
\label{eq:prl-conditional-state}
\end{equation}
By construction, $\rho_\lambda$ is a normalized density operator on
$V(\lambda)$ for every sector with $p_\lambda>0$.
Sectors with $p_\lambda=0$ make no contribution to the expressions
below and may therefore be omitted from all sector sums.

Define the sector mean vector
\begin{equation}
\bm\mu_\lambda
=
\left(
\Tr(\rho_\lambda X_1^{(\lambda)}),
\ldots,
\Tr(\rho_\lambda X_d^{(\lambda)})
\right),
\label{eq:sector-mean-vector}
\end{equation}
where $d=\dim\mathfrak g$, and define the global mean vector by
\begin{equation}
\overline{\bm\mu}
=
\sum_\lambda p_\lambda\bm\mu_\lambda .
\label{eq:global-mean-vector}
\end{equation}
With this notation in place, we arrive to our main result.

\begin{theorem}[Symmetry-resolved variance decomposition]
\label{thm:variance-decomposition}
Let $\rho$ be any density operator on
$\mathcal H_n=V^{\otimes n}$, with the isotypic decomposition
\eqref{eq:prl-isotypic}. Then the total variance of the collective
generators of $\mathfrak g$ admits the exact decomposition
\begin{equation}
\Delta_\rho^2(\mathfrak g)
=
\sum_\lambda p_\lambda
\Delta_{\rho_\lambda}^2(\mathfrak g)
+
\sum_\lambda p_\lambda
\left\|
\bm\mu_\lambda-\overline{\bm\mu}
\right\|^2.
\label{eq:prl-total-variance-decomposition}
\end{equation}
Equivalently, the second term can be written in the pairwise form
\begin{equation}
\sum_\lambda p_\lambda
\left\|
\bm\mu_\lambda-\overline{\bm\mu}
\right\|^2
=
\frac{1}{2}
\sum_{\lambda,\kappa}
p_\lambda p_\kappa
\left\|
\bm\mu_\lambda-\bm\mu_\kappa
\right\|^2.
\label{eq:prl-pairwise-dispersion}
\end{equation}
\end{theorem}

\begin{proof}
By Eq.~\eqref{eq:prl-block-action}, each collective generator is
block diagonal with respect to the isotypic decomposition,
\begin{equation}
X_a
=
\bigoplus_{\lambda\in\Lambda_n(V)}
\left(
I_{\mathcal M_\lambda}
\otimes X_a^{(\lambda)}
\right).
\end{equation}
The same is therefore true of its square,
\begin{equation}
X_a^2
=
\bigoplus_{\lambda\in\Lambda_n(V)}
\left(
I_{\mathcal M_\lambda}
\otimes
\bigl(X_a^{(\lambda)}\bigr)^2
\right).
\end{equation}
Consequently, both the first and second moments of $X_a$ depend only
on the diagonal isotypic blocks $P_\lambda\rho P_\lambda$ of the
state. In particular, coherences between distinct isotypic components
do not contribute to either moment.

Using Eqs.~\eqref{eq:prl-sector-weight} and
\eqref{eq:prl-conditional-state}, together with the fact that the
collective generators act trivially on the multiplicity spaces, we
obtain
\begin{align}
\Tr(\rho X_a)
&=
\sum_\lambda
p_\lambda
\Tr\!\left(
\rho_\lambda X_a^{(\lambda)}
\right),
\label{eq:sector-first-moment}
\\
\Tr(\rho X_a^2)
&=
\sum_\lambda
p_\lambda
\Tr\!\left[
\rho_\lambda
\bigl(X_a^{(\lambda)}\bigr)^2
\right].
\label{eq:sector-second-moment}
\end{align}
Hence
\begin{align}
\Var_\rho(X_a)
&=
\Tr(\rho X_a^2)
-
\bigl[\Tr(\rho X_a)\bigr]^2
\nonumber\\
&=
\sum_\lambda p_\lambda
\Tr\!\left[
\rho_\lambda
\bigl(X_a^{(\lambda)}\bigr)^2
\right]
-
\left[
\sum_\lambda p_\lambda
\Tr\!\left(
\rho_\lambda X_a^{(\lambda)}
\right)
\right]^2.
\label{eq:single-generator-variance}
\end{align}
Adding and subtracting
\[
\sum_\lambda p_\lambda
\left[
\Tr\!\left(
\rho_\lambda X_a^{(\lambda)}
\right)
\right]^2
\]
gives
\begin{align}
\Var_\rho(X_a)
&=
\sum_\lambda p_\lambda
\Var_{\rho_\lambda}
\!\left(X_a^{(\lambda)}\right)
\nonumber\\
&\quad+
\sum_\lambda p_\lambda
\left[
\Tr\!\left(
\rho_\lambda X_a^{(\lambda)}
\right)
\right]^2
-
\left[
\sum_\lambda p_\lambda
\Tr\!\left(
\rho_\lambda X_a^{(\lambda)}
\right)
\right]^2.
\label{eq:single-generator-decomposition}
\end{align}

Summing Eq.~\eqref{eq:single-generator-decomposition} over the
orthonormal generator basis $\{X_a\}_{a=1}^{d}$ yields
\begin{align}
\Delta_\rho^2(\mathfrak g)
&=
\sum_\lambda p_\lambda
\Delta_{\rho_\lambda}^2(\mathfrak g)
+
\sum_\lambda p_\lambda
\|\bm\mu_\lambda\|^2
-
\left\|
\sum_\lambda p_\lambda\bm\mu_\lambda
\right\|^2
\nonumber\\
&=
\sum_\lambda p_\lambda
\Delta_{\rho_\lambda}^2(\mathfrak g)
+
\sum_\lambda p_\lambda
\|\bm\mu_\lambda\|^2
-
\|\overline{\bm\mu}\|^2.
\label{eq:variance-intermediate}
\end{align}
Since $\sum_\lambda p_\lambda=1$, the last two terms satisfy
\begin{align}
\sum_\lambda p_\lambda
\|\bm\mu_\lambda\|^2
-
\|\overline{\bm\mu}\|^2
&=
\sum_\lambda p_\lambda
\left\|
\bm\mu_\lambda-\overline{\bm\mu}
\right\|^2.
\label{eq:mean-dispersion-identity}
\end{align}
Substitution into Eq.~\eqref{eq:variance-intermediate} proves
Eq.~\eqref{eq:prl-total-variance-decomposition}.

It remains to establish the equivalent pairwise form. Expanding the
squared norm gives
\begin{align}
\frac{1}{2}
\sum_{\lambda,\kappa}
p_\lambda p_\kappa
\left\|
\bm\mu_\lambda-\bm\mu_\kappa
\right\|^2
&=
\frac{1}{2}
\sum_{\lambda,\kappa}
p_\lambda p_\kappa
\left(
\|\bm\mu_\lambda\|^2
+
\|\bm\mu_\kappa\|^2
-
2\bm\mu_\lambda\!\cdot\!\bm\mu_\kappa
\right)
\nonumber\\
&=
\sum_\lambda p_\lambda
\|\bm\mu_\lambda\|^2
-
\left\|
\sum_\lambda p_\lambda\bm\mu_\lambda
\right\|^2
\nonumber\\
&=
\sum_\lambda p_\lambda
\left\|
\bm\mu_\lambda-\overline{\bm\mu}
\right\|^2,
\end{align}
where we used $\sum_\lambda p_\lambda=1$ and
Eq.~\eqref{eq:global-mean-vector}. This proves
Eq.~\eqref{eq:prl-pairwise-dispersion} and completes the proof.
\end{proof}

Theorem~\ref{thm:variance-decomposition} shows that the uncertainty of
a reducible representation contains two distinct contributions:
quantum fluctuations intrinsic to the individual irreducible sectors
and a nonnegative classical dispersion of their mean generator
vectors. In particular, intersector coherences do not affect the total
variance, since they do not enter the first or second moments of the
block-diagonal collective generators. Equation
\eqref{eq:prl-total-variance-decomposition} thus provides a
symmetry-resolved quantum analogue of the law of total variance.

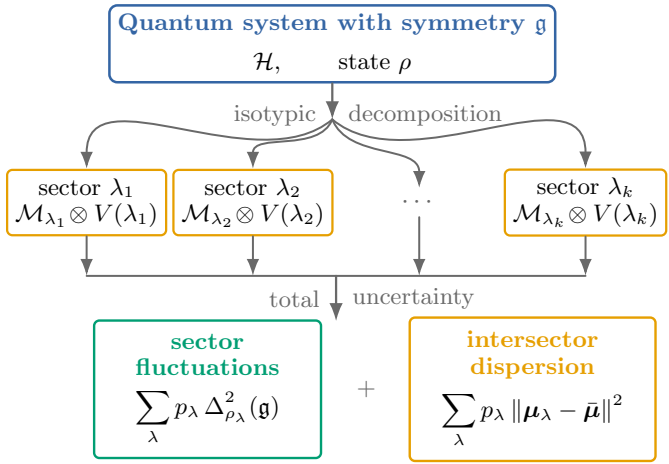
\begin{figure}[!h]

\centering

\begin{tikzpicture}[
    >=Latex,
    line cap=round,
    line join=round,
    font=\scriptsize,
sector/.style={
    draw=prlOrange,
    rounded corners=2pt,
    line width=0.85pt,
    minimum width=1.50cm,
    minimum height=0.82cm,
    align=center,
    inner xsep=3pt
},
    contribution/.style={
        rounded corners=2pt,
        line width=0.9pt,
        minimum height=1.12cm,
        align=center,
        inner xsep=5pt,
        inner ysep=5pt
    }
]


\definecolor{prlBlue}{RGB}{52,101,164}
\definecolor{prlOrange}{RGB}{230,159,0}
\definecolor{prlGreen}{RGB}{0,158,115}
\definecolor{prlGray}{RGB}{105,105,105}


\node[
    draw=prlBlue,
    rounded corners=3pt,
    line width=0.95pt,
    minimum width=4.55cm,
    minimum height=0.95cm,
    align=center
] (system) at (0,3.05)
{
    {\color{prlBlue}\bfseries Quantum system with symmetry $\mathfrak g$}
    \\[1.5mm]
    $\mathcal H$,
    \qquad state $\rho$
};


\coordinate (branchTop) at (0,2.05);

\draw[
    ->,
    line width=0.85pt,
    prlGray
]
(system.south) -- (branchTop);


\node[
    anchor=east,
    text=prlGray
] at (-0.10,2.11)
{
    isotypic
};

\node[
    anchor=west,
    text=prlGray
] at (0.10,2.11)
{
    decomposition
};


\node[sector] (s1) at (-3.25,0.95)
{
    sector $\lambda_1$
    \\[-0.2mm]
    $\mathcal M_{\lambda_1}\!\otimes V(\lambda_1)$
};

\node[sector] (s2) at (-1.05,0.95)
{
    sector $\lambda_2$
    \\[-0.2mm]
    $\mathcal M_{\lambda_2}\!\otimes V(\lambda_2)$
};

\node[
    font=\normalsize,
    text=prlGray
] (sdots) at (1.15,0.95)
{
    $\cdots$
};

\node[sector] (sk) at (3.35,0.95)
{
    sector $\lambda_k$
    \\[-0.2mm]
    $\mathcal M_{\lambda_k}\!\otimes V(\lambda_k)$
};


\draw[
    ->,
    line width=0.75pt,
    prlGray
]
(branchTop)
to[out=-140,in=90]
(s1.north);

\draw[
    ->,
    line width=0.75pt,
    prlGray
]
(branchTop)
to[out=-112,in=90]
(s2.north);

\draw[
    ->,
    line width=0.75pt,
    prlGray
]
(branchTop)
to[out=-68,in=90]
(sdots.north);

\draw[
    ->,
    line width=0.75pt,
    prlGray
]
(branchTop)
to[out=-40,in=90]
(sk.north);


\coordinate (c1) at (-3.25,-0.02);
\coordinate (c2) at (-1.05,-0.02);
\coordinate (c3) at (1.15,-0.02);
\coordinate (c4) at (3.35,-0.02);

\draw[
    ->,
    line width=0.75pt,
    prlGray
]
(s1.south) -- (c1);

\draw[
    ->,
    line width=0.75pt,
    prlGray
]
(s2.south) -- (c2);

\draw[
    ->,
    line width=0.75pt,
    prlGray
]
(sdots.south) -- (c3);

\draw[
    ->,
    line width=0.75pt,
    prlGray
]
(sk.south) -- (c4);


\draw[
    line width=0.75pt,
    prlGray
]
(c1) -- (c4);

\coordinate (merge) at (0.05,-0.02);


\draw[
    ->,
    line width=0.85pt,
    prlGray
]
(merge) -- (0.05,-0.62);


\node[
    anchor=east,
    text=prlGray
] at (-0.05,-0.32)
{
    total
};

\node[
    anchor=west,
    text=prlGray
] at (0.15,-0.32)
{
    uncertainty
};


\node[
    contribution,
    draw=prlGreen,
    text width=2.65cm
] (intrinsic) at (-1.65,-1.52)
{
    {\color{prlGreen}\bfseries sector}
    \\[-0.2mm]
    {\color{prlGreen}\bfseries fluctuations}
    \\[1.4mm]
    $\displaystyle
    \sum_{\lambda}
    p_\lambda\,
    \Delta_{\rho_\lambda}^{2}(\mathfrak g)$
};


\node[
    font=\small\bfseries,
    text=prlGray
] at (0.45,-1.52)
{
    $+$
};


\node[
    contribution,
    draw=prlOrange,
    text width=2.90cm
] (dispersion) at (2.65,-1.52)
{
    {\color{prlOrange}\bfseries intersector}
    \\[-0.2mm]
    {\color{prlOrange}\bfseries dispersion}
    \\[1.4mm]
    $\displaystyle
    \sum_{\lambda}
    p_\lambda
    \left\|
    \bm{\mu}_\lambda-\overline{\bm{\mu}}
    \right\|^2$
};

\end{tikzpicture}

\caption{
Symmetry-resolved structure of collective uncertainty.
After isotypic decomposition, the total 
variance separates exactly into
intrinsic quantum fluctuations and nonnegative
dispersion between sectors.
}

\label{fig:symmetry-decomposition}

\end{figure}

The exact decomposition converts irreducible uncertainty relations into
sharp bounds for arbitrary reducible representations. For an irreducible
sector $V(\lambda)$, the pure-state uncertainty relation
\begin{equation}
\Delta_{|\psi\rangle}^2(\mathfrak g)
\geq
c(\lambda)
=
2\langle\lambda,\delta\rangle
\label{eq:pure-irrep-bound}
\end{equation}
holds for every normalized pure state
$|\psi\rangle\in V(\lambda)$
\cite{dammeier2015,deGuise2018,HuJingZhangXiao2026}.
The same lower bound extends to arbitrary mixed states, as follows from
the concavity of the variance.

\begin{proposition}[Mixed-state irreducible uncertainty bound]
\label{prop:mixed-irrep-bound}
Let $\rho$ be any density operator on an irreducible highest-weight
representation $V(\lambda)$ of the compact semisimple Lie algebra
$\mathfrak g$. Then
\begin{equation}
\Delta_{\rho}^2(\mathfrak g)
\geq
c(\lambda)
=
2\langle\lambda,\delta\rangle .
\label{eq:prl-irrep-bound}
\end{equation}
\end{proposition}

\begin{proof}
Let
\begin{equation}
\rho
=
\sum_i q_i
|\psi_i\rangle\langle\psi_i|,
\qquad
q_i\geq0,
\qquad
\sum_i q_i=1,
\label{eq:mixed-state-decomposition}
\end{equation}
be any pure-state ensemble decomposition of $\rho$, with
$|\psi_i\rangle\in V(\lambda)$. For any Hermitian observable $A$,
linearity of the trace gives
\begin{align}
\Var_\rho(A)
&=
\Tr(\rho A^2)
-
\bigl[\Tr(\rho A)\bigr]^2
\nonumber\\
&=
\sum_i q_i
\langle\psi_i|A^2|\psi_i\rangle
-
\left(
\sum_i q_i
\langle\psi_i|A|\psi_i\rangle
\right)^2.
\label{eq:mixed-variance-start}
\end{align}
Adding and subtracting
\[
\sum_i q_i
\langle\psi_i|A|\psi_i\rangle^2
\]
yields
\begin{align}
\Var_\rho(A)
&=
\sum_i q_i
\Var_{\psi_i}(A)
+
\sum_i q_i
\langle\psi_i|A|\psi_i\rangle^2
-
\left(
\sum_i q_i
\langle\psi_i|A|\psi_i\rangle
\right)^2
\nonumber\\
&\geq
\sum_i q_i
\Var_{\psi_i}(A),
\label{eq:variance-concavity}
\end{align}
because the last two terms form the nonnegative classical variance of
the expectation values
$\langle\psi_i|A|\psi_i\rangle$ with probabilities $q_i$.

Applying Eq.~\eqref{eq:variance-concavity} to each generator
$X_a^{(\lambda)}$ and summing over the orthonormal generator basis gives
\begin{align}
\Delta_{\rho}^2(\mathfrak g)
&=
\sum_{a=1}^{d}
\Var_\rho\!\left(X_a^{(\lambda)}\right)
\nonumber\\
&\geq
\sum_i q_i
\sum_{a=1}^{d}
\Var_{\psi_i}\!\left(X_a^{(\lambda)}\right)
\nonumber\\
&=
\sum_i q_i
\Delta_{|\psi_i\rangle}^2(\mathfrak g).
\label{eq:mixed-total-variance-concavity}
\end{align}
Using the pure-state irreducible bound
\eqref{eq:pure-irrep-bound} for every $|\psi_i\rangle$ therefore gives
\begin{align}
\Delta_{\rho}^2(\mathfrak g)
&\geq
\sum_i q_i\,c(\lambda)
\nonumber\\
&=
c(\lambda)
=
2\langle\lambda,\delta\rangle,
\end{align}
where $\sum_i q_i=1$. This proves
Eq.~\eqref{eq:prl-irrep-bound}.
\end{proof}

Thus, the irreducible uncertainty constant
$c(\lambda)=2\langle\lambda,\delta\rangle$ bounds the total variance
for every density operator on $V(\lambda)$, not only for pure states.
In particular, it applies to each conditional sector state
$\rho_\lambda$ appearing in
Theorem~\ref{thm:variance-decomposition}. Since the intersector
dispersion term in Eq.~\eqref{eq:prl-total-variance-decomposition} is
nonnegative, combining the proposition with the exact variance
decomposition yields our central sector-resolved uncertainty relation,
\begin{equation}
\Delta_\rho^2(\mathfrak g)
\geq
\sum_{\lambda}
p_\lambda c(\lambda)
=
\sum_{\lambda}
p_\lambda\,2\langle\lambda,\delta\rangle .
\label{eq:prl-sector-bound}
\end{equation}

\begin{figure}[!h]
\centering

\begin{tikzpicture}[
    >=Latex,
    line cap=round,
    line join=round,
    every node/.style={font=\scriptsize}
]


\definecolor{sectorBlue}{RGB}{70,120,190}
\definecolor{minimumOrange}{RGB}{230,159,0}
\definecolor{textGrey}{RGB}{90,90,90}

\definecolor{planeFill}{RGB}{232,238,244}


\coordinate (O)  at (-2.30,-1.75);
\coordinate (X)  at ( 2.95,-1.75);
\coordinate (Y)  at (-0.65,-0.30);
\coordinate (XY) at ( 4.60,-0.30);

\coordinate (Zt) at (-2.30,2.55);


\fill[
    planeFill
]
(O)--(X)--(XY)--(Y)--cycle;


\draw[
    ->,
    textGrey,
    line width=0.95pt
]
(O)--(Zt)
node[
    above,
    align=center
]
{
    $c(\lambda)
    =2\langle\lambda,\delta\rangle$
};


\node[
    text=textGrey,
    font=\scriptsize\itshape
] at (1.45,-1.66)
{
    sector space
};

\node[
    anchor=east,
    text=textGrey
] at (-2.43,-1.75)
{
    $0$
};

%

\coordinate (A) at (-1.25,-1.30);
\coordinate (B) at ( 0.55,-0.92);
\coordinate (S) at ( 2.25,-1.08);
\coordinate (C) at (-0.35,-0.58);


\coordinate (At) at ($(A)+(0,1.78)$);
\coordinate (Bt) at ($(B)+(0,2.12)$);
\coordinate (St) at ($(S)+(0,1.30)$);
\coordinate (Ct) at ($(C)+(0,2.55)$);


\fill[
    sectorBlue
]
(A) circle (1.35pt);

\fill[
    sectorBlue
]
(B) circle (1.35pt);

\fill[
    minimumOrange
]
(S) circle (1.50pt);

\fill[
    sectorBlue
]
(C) circle (1.35pt);


\node[
    anchor=north,
    text=textGrey
] at ($(A)+(0,-0.11)$)
{
    $\lambda_A$
};

\node[
    anchor=north,
    text=textGrey
] at ($(B)+(0,-0.11)$)
{
    $\lambda_B$
};

\node[
    anchor=north,
    text=textGrey
] at ($(S)+(0,-0.11)$)
{
    $\lambda_\ast$
};

\node[
    anchor=north,
    text=textGrey
] at ($(C)+(0,-0.11)$)
{
    $\lambda_C$
};

\node[
    text=textGrey,
    font=\normalsize
] at (1.55,-0.72)
{
    $\cdots$
};


\draw[
    sectorBlue,
    line width=1.05pt
]
(A)--(At);

\fill[
    sectorBlue
]
(At) circle (2.25pt);

\node[
    anchor=south,
    text=sectorBlue
] at ($(At)+(0,0.11)$)
{
    $c(\lambda_A)$
};


\draw[
    sectorBlue,
    line width=1.05pt
]
(B)--(Bt);

\fill[
    sectorBlue
]
(Bt) circle (2.25pt);

\node[
    anchor=south,
    text=sectorBlue
] at ($(Bt)+(0,0.11)$)
{
    $c(\lambda_B)$
};


\draw[
    minimumOrange,
    line width=1.30pt
]
(S)--(St);

\fill[
    minimumOrange
]
(St) circle (2.85pt);

\node[
    anchor=south,
    text=minimumOrange
] at ($(St)+(0,0.12)$)
{
    $c(\lambda_\ast)=C_{\min}$
};

\node[
    anchor=west,
    text=minimumOrange,
    align=left
] at ($(S)+(0.20,0.62)$)
{
    lowest-uncertainty\\[-0.4mm]
    sector
};


\draw[
    sectorBlue,
    line width=1.05pt
]
(C)--(Ct);

\fill[
    sectorBlue
]
(Ct) circle (2.25pt);

\node[
    anchor=south,
    text=sectorBlue
] at ($(Ct)+(0,0.11)$)
{
    $c(\lambda_C)$
};


\node[
    align=center,
    text=textGrey
] at (2.85,2.35)
{
    sharp floor selected by the\\[-0.4mm]
    lowest accessible sector
};

\draw[
    ->,
    textGrey,
    line width=0.70pt
]
(2.70,2.10)
to[out=-105,in=65]
($(St)+(0.05,0.06)$);

\end{tikzpicture}

\caption{
Sector-resolved uncertainty landscape for a reducible representation.
Each irreducible sector $\lambda$ is represented by a point in the
sector space, while the vertical direction gives its uncertainty level
$c(\lambda)=2\langle\lambda,\delta\rangle$.
The lowest accessible sector $\lambda_\ast$ determines the sharp
state-independent uncertainty floor
$C_{\min}=c(\lambda_\ast)$.
}

\label{fig:sector-landscape}

\end{figure}
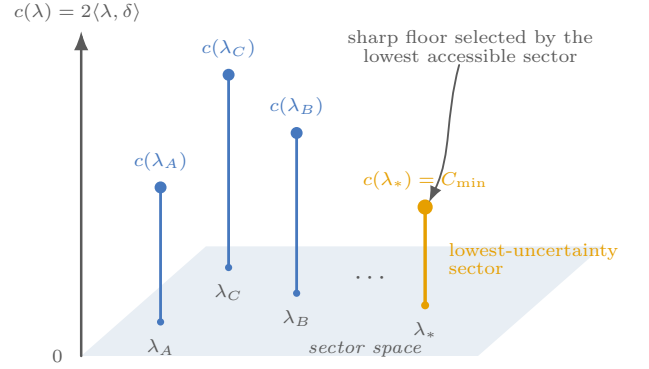
This relation provides more than the global minimum uncertainty: it
resolves the uncertainty scale according to the symmetry-sector
composition of an arbitrary quantum state.  Since the right-hand side
is a convex combination of irreducible uncertainty constants, the
minimum uncertainty over the full reducible representation is
determined entirely by the lowest-uncertainty sector,
\begin{equation}
\min_\rho
\Delta_\rho^2(\mathfrak g)
=
C_{\min}(\mathcal H_n)
=
\min_{\lambda\in\Lambda_n(V)}
2\langle\lambda,\delta\rangle .
\label{eq:prl-sharp-bound}
\end{equation}

The bound is attainable by a generalized coherent state in any
irreducible sector minimizing
$2\langle\lambda,\delta\rangle$, embedded into the corresponding
isotypic component
\cite{Delbourgo1977,DelbourgoFox1977,Perelomov1986}.
Therefore, the uncertainty scale of a reducible quantum system is not
set by the Hilbert-space dimension, but by the lowest-uncertainty symmetry
sector permitted by its representation structure.

This result establishes a general principle for collective quantum
systems: irreducible uncertainty relations extend to arbitrary
multipartite representations through an exact decomposition into
sector-resolved quantum fluctuations and symmetry-induced dispersion.

\section{Application to collective spin systems}

Collective spin systems provide a direct illustration of how symmetry
sectors determine state-independent uncertainty.  We first consider
$n$ spin-$1/2$ particles, whose collective Hilbert space is resolved by
the Clebsch--Gordan decomposition
\cite{FultonHarris,Hall2015,Hayashi2017},
\begin{equation}
(\mathbb C^2)^{\otimes n}
\cong
\bigoplus_{j\in\mathcal J_n}
\left(
\mathcal M_{n,j}\otimes V^{(j)}
\right),
\label{eq:spin-half-prl-decomposition}
\end{equation}
where $V^{(j)}$ is the spin $j$ irreducible representation and
\begin{equation}
\mathcal J_n
=
\left\{
\frac n2,\frac n2-1,\ldots,j_{\min}
\right\},
j_{\min}
=
\begin{cases}
\frac12,& n\ {\rm odd},\\
0,& n\ {\rm even}.
\end{cases}
\label{eq:spin-half-prl-jmin}
\end{equation}
For $\mathfrak{su}(2)$, the irreducible uncertainty constant reduces
to
\[
c(j)=j
\]
\cite{dammeier2015,deGuise2018,HuJingZhangXiao2026}.
Therefore, the sector-resolved uncertainty relation becomes
\begin{equation}
\Delta_\rho^2(\mathfrak{su}(2))
\geq
\sum_j p_j j
\geq
\begin{cases}
\frac12,& n\ {\rm odd},\\[1mm]
0,& n\ {\rm even}.
\end{cases}
\label{eq:spin-half-prl-bound}
\end{equation}
This result rigorously proves the parity-dependent uncertainty
behavior conjectured in Ref.~\cite{Wang2026}: odd numbers of spin-$1/2$
particles possess a non-zero uncertainty floor, whereas even numbers can
access a zero-uncertainty symmetry sector.

\begin{figure}[!h]
\centering

\begin{tikzpicture}[
    >=Latex,
    line cap=round,
    line join=round,
    every node/.style={font=\scriptsize}
]


\definecolor{oddBlue}{RGB}{70,120,190}
\definecolor{evenOrange}{RGB}{230,159,0}
\definecolor{textGrey}{RGB}{90,90,90}
\definecolor{lightGrey}{RGB}{175,175,175}


\node[
    anchor=west,
    font=\scriptsize\bfseries
] at (1.45,5.95)
{
    (a)
};

\node[
    anchor=west,
    font=\scriptsize\bfseries,
    text=textGrey
] at (1.90,5.95)
{
    sector mechanism
};


\node[
    font=\scriptsize\bfseries,
    text=oddBlue
] at (1.45,5.42)
{
    odd $n$
};

\draw[
    line width=0.85pt,
    textGrey
]
(0.90,4.95)--(1.75,4.95);

\node[
    anchor=east,
    text=textGrey
] at (0.77,4.95)
{
    $j=\frac{n}{2}$
};

\node[
    text=textGrey,
    font=\normalsize
] at (1.33,4.40)
{
    $\vdots$
};

\draw[
    line width=1.35pt,
    oddBlue
]
(0.90,3.75)--(1.75,3.75);

\node[
    anchor=east,
    text=oddBlue
] at (0.77,3.75)
{
    $j_{\min}=\frac{1}{2}$
};

\node[
    text=oddBlue,
    align=center
] at (1.33,3.25)
{
    no singlet
};


\node[
    font=\scriptsize\bfseries,
    text=evenOrange
] at (4.05,5.42)
{
    even $n$
};

\draw[
    line width=0.85pt,
    textGrey
]
(3.45,4.95)--(4.30,4.95);

\node[
    anchor=west,
    text=textGrey
] at (4.45,4.95)
{
    $j=\frac{n}{2}$
};

\node[
    text=textGrey,
    font=\normalsize
] at (3.88,4.48)
{
    $\vdots$
};

\draw[
    line width=0.85pt,
    textGrey
]
(3.45,4.10)--(4.30,4.10);

\node[
    anchor=west,
    text=textGrey
] at (4.45,4.10)
{
    $j=1$
};

\draw[
    line width=1.35pt,
    evenOrange
]
(3.45,3.55)--(4.30,3.55);

\node[
    anchor=west,
    text=evenOrange
] at (4.45,3.55)
{
    $j_{\min}=0$
};

\node[
    text=evenOrange,
    align=center
] at (3.88,3.08)
{
    singlet
};


\draw[
    densely dotted,
    lightGrey,
    line width=0.45pt
]
(0.45,2.82)--(5.60,2.82);


\node[
    anchor=west,
    font=\scriptsize\bfseries
] at (1.55,2.55)
{
    (b)
};

\node[
    anchor=west,
    font=\scriptsize\bfseries,
    text=textGrey
] at (2.00,2.55)
{
    parity outcome
};

%
%

\draw[
    ->,
    line width=0.85pt,
    textGrey
]
(0.92,0)--(5.60,0)
node[
    right,
    text=textGrey
]
{
    $n$
};

\draw[
    ->,
    line width=0.85pt,
    textGrey
]
(0.92,0)--(0.92,2.32);

\node[
    anchor=east,
    text=textGrey
] at (0.78,2.28)
{
    $C_{\min}$
};


\node[
    anchor=east,
    text=textGrey
] at (0.72,0)
{
    $0$
};

\node[
    anchor=east,
    text=textGrey
] at (0.72,1.00)
{
    $\frac{1}{2}$
};

\node[
    anchor=east,
    text=textGrey
] at (0.72,2.00)
{
    $1$
};


\draw[
    dashed,
    line width=0.70pt,
    oddBlue
]
(0.92,1.00)--(4.98,1.00);

\node[
    anchor=west,
    text=oddBlue
] at (5.05,1.00)
{
    $C_{\min}=\frac{1}{2}$
};


\draw[
    dash pattern=on 1.2pt off 2.4pt,
    line width=0.75pt,
    evenOrange
]
(0.92,2.00)--(4.98,2.00);

\node[
    anchor=west,
    text=evenOrange
] at (5.05,2.00)
{
    $C_{\min}^{\rm red}=1$
};


\node[
    anchor=west,
    text=evenOrange
] at (5.05,0.18)
{
    $C_{\min}=0$
};


\foreach \x/\lab in {
    1.35/1,
    1.82/2,
    2.29/3,
    2.76/4,
    3.23/5,
    3.70/6,
    4.17/7,
    4.64/8
}
{
    \draw[
        line width=0.50pt,
        textGrey
    ]
    (\x,0.055)--(\x,-0.055);

    \node[
        anchor=north,
        text=textGrey
    ] at (\x,-0.13)
    {
        $\lab$
    };
}


\fill[oddBlue]
(1.35,1.00)
circle(2.0pt);

\fill[oddBlue]
(2.29,1.00)
circle(2.0pt);

\fill[oddBlue]
(3.23,1.00)
circle(2.0pt);

\fill[oddBlue]
(4.17,1.00)
circle(2.0pt);


\fill[evenOrange]
(1.77,-0.055)
rectangle
(1.87,0.055);

\fill[evenOrange]
(2.71,-0.055)
rectangle
(2.81,0.055);

\fill[evenOrange]
(3.65,-0.055)
rectangle
(3.75,0.055);

\fill[evenOrange]
(4.59,-0.055)
rectangle
(4.69,0.055);


\draw[
    evenOrange,
    line width=0.95pt,
    fill=white
]
(1.77,1.95)
rectangle
(1.87,2.05);

\draw[
    evenOrange,
    line width=0.95pt,
    fill=white
]
(2.71,1.95)
rectangle
(2.81,2.05);

\draw[
    evenOrange,
    line width=0.95pt,
    fill=white
]
(3.65,1.95)
rectangle
(3.75,2.05);

\draw[
    evenOrange,
    line width=0.95pt,
    fill=white
]
(4.59,1.95)
rectangle
(4.69,2.05);


\node[
    anchor=west,
    text=oddBlue,
    font=\scriptsize\bfseries
] at (1.10,1.27)
{
    odd $n$
};

\node[
    anchor=west,
    text=evenOrange,
    font=\scriptsize\bfseries
] at (1.10,0.28)
{
    even $n$
};

\node[
    anchor=west,
    text=evenOrange,
    font=\scriptsize
] at (1.10,2.24)
{
    even $n$, reduced
};

\end{tikzpicture}

\caption{
Parity-controlled uncertainty in multipartite spin-$1/2$ systems, where $n$ denotes the number of particles. 
(a) For odd $n$, the lowest accessible sector is $j_{\min}=1/2$,
whereas even $n$ admits the singlet sector $j_{\min}=0$.
(b) Accordingly, the full-space uncertainty floor is
$C_{\min}=1/2$ for odd $n$ and $C_{\min}=0$ for even $n$. 
In the latter case, the reduced-space uncertainty
floor is $C_{\min}^{\rm red}=1$.
}

\label{fig:spin-half-parity}

\end{figure}
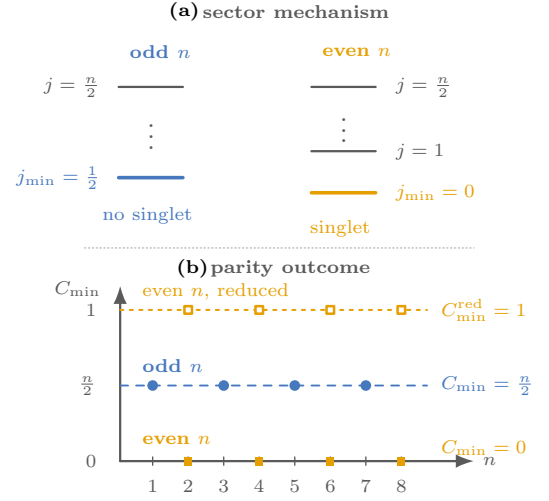

The parity dependence is thus a direct consequence of the available
irreducible sectors. For odd numbers of spin-$1/2$ particles, the
Clebsch--Gordan decomposition contains no trivial representation, and
the lowest-uncertainty sector has spin $j=1/2$. A finite uncertainty floor
therefore remains for every state. In contrast, even numbers of particles
admit a singlet sector $V^{(0)}$, on which all collective generators
vanish identically. Consequently, the zero uncertainty value
\[
\Delta_\rho^2(\mathfrak{su}(2))=0
\]
is attainable when the state is supported entirely on the singlet subspace. This representation-theoretic mechanism explains why the uncertainty floor disappears without requiring any special state preparation.

The same framework also applies when symmetry sectors are removed or
restricted. For example, excluding the singlet sector of two qubits
leaves the triplet representation
\[
(\mathbb C^2)^{\otimes2}_{\rm red}=V^{(1)},
\]
and the sharp uncertainty bound on the reduced space becomes
\begin{equation}
\Delta_\rho^2(\mathfrak{su}(2))_{red}\geq1 .
\label{eq:triplet-bound-prl}
\end{equation}
This bound holds for arbitrary mixed states in the reduced space and is
saturated by spin-coherent states in the $j=1$ sector
\cite{Perelomov1986,ZhangFengGilmore1990}.

The mechanism is not restricted to qubit systems and extends naturally
to higher-spin representations, where irreducible sectors may occur
with nontrivial multiplicities.  For two spin-$1$ particles, for
example,
\begin{equation}
V^{(1)}\otimes V^{(1)}
\cong
V^{(2)}
\oplus
V^{(1)}
\oplus
V^{(0)},
\end{equation}
so that the singlet already provides a zero-level symmetry sector and
hence $C_{\min}=0$.  For three spin-$1$ particles,
\begin{equation}
(V^{(1)})^{\otimes3}
\cong
V^{(3)}
\oplus
2V^{(2)}
\oplus
3V^{(1)}
\oplus
V^{(0)},
\end{equation}
demonstrating explicitly the appearance of multiplicity spaces in the
reducible tensor-product representation.

More generally, write
\begin{equation}\label{eq:spin1-general-decomposition}
(V^{(1)})^{\otimes n}
\cong
\bigoplus_{j=0}^{n}
m_{n,j}V^{(j)},
\end{equation}
where $m_{n,j}$ is the multiplicity of the irreducible spin-$j$
representation $V^{(j)}$. Equivalently, in the notation of the
isotypic decomposition,
\begin{equation}
m_{n,j}
=
\dim\mathcal M_{n,j},
\end{equation}
where $\mathcal M_{n,j}$ is the corresponding multiplicity space.
For the uncertainty floor, only the two lowest spin sectors are needed.
Their multiplicities can be determined explicitly from the character
of the spin-$1$ representation.

\begin{proposition}[Lowest spin-sector multiplicities]
\label{prop:spin1-multiplicities}
For the $n$-fold tensor product of the spin-$1$ representation of
$\mathfrak{su}(2)$,
\begin{equation}
(V^{(1)})^{\otimes n}
\cong
\bigoplus_{j=0}^{n}
m_{n,j}V^{(j)},
\end{equation}
the multiplicity of the spin-$j$ sector is given by
\begin{equation}
m_{n,j}
=
a_{n,j}-a_{n,j+1},
\label{eq:spin1-multiplicity-relation}
\end{equation}
where
\begin{equation}
a_{n,j}
=
\sum_{r=0}^{\lfloor (n-j)/2\rfloor}
\frac{n!}
{r!(r+j)!(n-2r-j)!}.
\label{eq:spin1-weight-multiplicity}
\end{equation}
In particular,
\begin{equation}
m_{n,0}
=
\sum_{r=0}^{\lfloor n/2\rfloor}
\frac{n!(3r+1-n)}
{r!(r+1)!(n-2r)!},
\label{eq:spin1-mn0}
\end{equation}
and
\begin{equation}
m_{n+1,0}=m_{n,1}.
\label{eq:spin1-mn1-mn0}
\end{equation}
Consequently,
\begin{equation}
m_{n,0}>0
\quad (n\geq2),
\qquad
m_{n,1}>0
\quad (n\geq1).
\label{eq:spin1-lowest-multiplicities}
\end{equation}
\end{proposition}

\begin{proof}
The character of the irreducible spin-$j$ representation \cite{FultonHarris} is 
\begin{equation}
\chi_j(x)
=
\sum_{k=-j}^{j}x^k
\label{eq:spin1-character}
\end{equation}
 In particular, the character of the spin-$1$
representation is
\begin{equation}
\chi_1(x)
=
x^{-1}+1+x,
\end{equation}
and therefore
\begin{equation}
\chi_{(V^{(1)})^{\otimes n}}(x)
=
(x^{-1}+1+x)^n.
\label{eq:spin1-tensor-character}
\end{equation}

Let $a_{n,j}$ denote the multiplicity of weight $j$ in the tensor
product $(V^{(1)})^{\otimes n}$, equivalently the coefficient of
$x^j$ in Eq.~\eqref{eq:spin1-tensor-character}. To obtain a term
$x^j$, choose $r$ of the $n$ factors to contribute $x^{-1}$,
$r+j$ factors to contribute $x$, and the remaining
$n-2r-j$ factors to contribute $1$. Hence
\begin{align}
a_{n,j}
&=
[x^j](x^{-1}+1+x)^n
\nonumber\\
&=
\sum_{r=0}^{\lfloor (n-j)/2\rfloor}
\frac{n!}
{r!(r+j)!(n-2r-j)!},
\end{align}
which proves Eq.~\eqref{eq:spin1-weight-multiplicity}.

On the other hand, taking characters in
Eq.~\eqref{eq:spin1-general-decomposition} gives
\begin{equation}
(x^{-1}+1+x)^n
=
\sum_{J=0}^{n}
m_{n,J}\chi_J(x).
\label{eq:spin1-character-decomposition}
\end{equation}
For $j\geq0$, the coefficient of $x^j$ in $\chi_J(x)$ is equal to
one when $J\geq j$ and zero when $J<j$. It follows that
\begin{equation}
a_{n,j}
=
\sum_{J=j}^{n}m_{n,J}
=
m_{n,j}+m_{n,j+1}+\cdots+m_{n,n}.
\label{eq:spin1-weight-sector-relation}
\end{equation}
Taking the difference of Eq.~\eqref{eq:spin1-weight-sector-relation}
for consecutive weights therefore yields
\begin{equation}
m_{n,j}
=
a_{n,j}-a_{n,j+1},
\end{equation}
which proves Eq.~\eqref{eq:spin1-multiplicity-relation}.

For $j=0$, Eqs.~\eqref{eq:spin1-multiplicity-relation} and
\eqref{eq:spin1-weight-multiplicity} give
\begin{align}
m_{n,0}
&=
\sum_{r=0}^{\lfloor n/2\rfloor}
\frac{n!}{(r!)^2(n-2r)!}
\left(
1-\frac{n-2r}{r+1}
\right)
\nonumber\\
&=
\sum_{r=0}^{\lfloor n/2\rfloor}
\frac{n!(3r+1-n)}
{r!(r+1)!(n-2r)!},
\end{align}
which is Eq.~\eqref{eq:spin1-mn0}.

The relation between the two lowest multiplicities follows directly
from the Clebsch--Gordan decomposition. For $j\geq1$,
\begin{equation}
V^{(j)}\otimes V^{(1)}
\cong
V^{(j+1)}
\oplus
V^{(j)}
\oplus
V^{(j-1)}.
\label{eq:spin1-CG}
\end{equation}
A spin-$0$ representation in
$(V^{(1)})^{\otimes(n+1)}$ can therefore arise only by coupling a
spin-$1$ sector of $(V^{(1)})^{\otimes n}$ with the additional
spin-$1$ representation. Since
\begin{equation}
V^{(1)}\otimes V^{(1)}
\cong
V^{(2)}
\oplus
V^{(1)}
\oplus
V^{(0)}
\end{equation}
contains the singlet exactly once, each copy of $V^{(1)}$ at level
$n$ contributes exactly one singlet at level $n+1$. Thus
\begin{equation}
m_{n+1,0}=m_{n,1},
\end{equation}
which proves Eq.~\eqref{eq:spin1-mn1-mn0}.

Finally, since $m_{1,1}=1$ and each copy of $V^{(1)}$ at level $n$
contributes one copy of $V^{(1)}$ at level $n+1$ through
Eq.~\eqref{eq:spin1-CG}, induction gives
\begin{equation}
m_{n,1}>0,
\qquad
n\geq1.
\end{equation}
Since $m_{1,0}=0$, Eq.~\eqref{eq:spin1-mn1-mn0} then implies
\begin{equation}
m_{n,0}>0,
\qquad
n\geq2.
\end{equation}
This establishes Eq.~\eqref{eq:spin1-lowest-multiplicities}.
\end{proof}

The proposition shows in particular that the singlet sector occurs
for every $n\geq2$, while the spin-$1$ sector occurs for every
$n\geq1$. These are precisely the lowest-sector properties needed to
determine the sharp uncertainty floor.

The singlet is absent for a single spin, $m_{1,0}=0$, while the next ten multiplicities $m_{n, 0}, 2\leq n\leq 11$ are
\begin{equation}
1,1,3,6,15,36,91,232,603,1585.
\label{eq:spin1-singlet-values}
\end{equation}

Thus every tensor product of two or more spin-$1$ representations
contains a singlet sector. Since $c(j)=j$ for $\mathfrak{su}(2)$,
for $n\geq2$ we obtain
\begin{equation}
\label{eq:spin1-global-minimum}
j_{\min}
\equiv
\min_{j:m_{n,j}\neq0}j
=
0,
\qquad
C_{\min}
=
0.
\end{equation}
Moreover, Eq.~\eqref{eq:spin1-lowest-multiplicities} shows that the
spin-$1$ sector is present for every $n\geq1$. Hence, if the singlet
sector is excluded while the remaining tensor-product sectors are
retained, the lowest surviving sector is $j=1$, giving the sharp
uncertainty floor on the reduced space,
\begin{equation}
C_{\min}^{(\mathrm{red})}
=
1.
\label{eq:spin1-reduced-minimum}
\end{equation}

Thus, beyond spin-$1/2$, the state-independent uncertainty floor is
fixed by the lowest irreducible sector admitted by the
representation, rather than by the dimension of the underlying
Hilbert space.  Multiplicity determines how often a sector occurs, but
not the uncertainty associated with that sector.

\section{Conclusion}
We have established a symmetry-resolved framework for state-independent uncertainty in reducible representations.
The exact decomposition of the total variance into intrinsic fluctuations within irreducible sectors and nonnegative intersector dispersion shows that the uncertainty scale of a reducible quantum system is fixed by its representation structure rather than by Hilbert-space dimension.

The exact separation from multiplicity spaces from irreducible sectors also connects this framework with broader symmetry-based structures in
quantum information, including noiseless subsystems and symmetry-protected quantum processing.
Beyond uncertainty relations, the representation-theoretic characterization viewpoints may provide insights into symmetry-controlled quantum sensing and metrology, where accessible sectors can limit
achievable performance.

More generally, the present approach provides a framework to determine state-independent fluctuation bounds in quantum systems with more general symmetry structures, where the interplay between irreducible sectors and collective observables remains largely unexplored.

\bigskip
\centerline{\bf Acknowledgments}
The project is partially supported by the Simons Foundation grant no. MP-TSM-00002518.

\vskip 20pt

\noindent{\bf Data Availability Statement.}
All data of this work are included in the manuscript and its supplemental materials.

\vskip 20pt

\appendix

\end{document}